\documentclass[10pt, aps, twocolumn, pra, superscriptaddress]{revtex4-2}

\usepackage{multirow}
\usepackage{enumitem}
\usepackage{booktabs}
\usepackage{mathtools}
\usepackage{graphicx}
\usepackage[per-mode = symbol, bracket-unit-denominator = false]{siunitx}
\usepackage{braket}
\usepackage[switch]{lineno}
\usepackage[hidelinks]{hyperref}

\usepackage{amsfonts}

\begin{document}

\title{A chip-scale atomic beam source for non-classical light}

\author{Braden J.~Larsen}
\affiliation{JILA, University of Colorado, Boulder, CO, USA.}
\affiliation{National Institute of Standards and Technology, Boulder, CO, USA.}
\affiliation{Department of Physics, University of Colorado, Boulder, CO, USA.}
\affiliation{These authors contributed equally to this work.}
\author{Hagan Hensley}
\affiliation{JILA, University of Colorado, Boulder, CO, USA.}
\affiliation{National Institute of Standards and Technology, Boulder, CO, USA.}
\affiliation{Department of Physics, University of Colorado, Boulder, CO, USA.}
\affiliation{These authors contributed equally to this work.}
\author{Gabriela D.~Martinez}
\affiliation{National Institute of Standards and Technology, Boulder, CO, USA.}
\affiliation{Department of Physics, University of Colorado, Boulder, CO, USA.}
\author{Alexander Staron}
\affiliation{National Institute of Standards and Technology, Boulder, CO, USA.}
\affiliation{Department of Physics, University of Colorado, Boulder, CO, USA.}
\author{William R.~McGehee}
\affiliation{National Institute of Standards and Technology, Boulder, CO, USA.}
\author{John Kitching}
\affiliation{National Institute of Standards and Technology, Boulder, CO, USA.}
\author{James K.~Thompson}
\affiliation{JILA, University of Colorado, Boulder, CO, USA.}
\affiliation{National Institute of Standards and Technology, Boulder, CO, USA.}
\affiliation{Department of Physics, University of Colorado, Boulder, CO, USA.}
\date{\today}

\begin{abstract}
Room temperature thermal atoms have proven to be a powerful resource for magnetometry, electrometry, atom-entanglement generation, and robust atomic clocks.  Recent efforts have sought to realize compact and highly manufacturable atomic vapors and atomic beams for chip-scale magnetometry and atomic clocks.  Here, we show that a chip-scale rubidium beam source can be integrated with a high finesse cavity-QED system to generate non-classical light.  By demonstrating the compatibility of these two technologies, we open a new path for distributed sources of non-classical light and set the stage for using cavity-QED to enhance the performance of chip-scale magnetometers and atomic clocks.
\end{abstract}


\maketitle

\section*{Introduction}

Cavity-QED experiments with atoms often rely on laser cooling and trapping atoms at microKelvin temperature in order to hold them within the cavity mode \cite{Cline2024, Young2024, Thomas2022, Thompson2013,Fortier2007, Birnbaum2005, Deist2022,Dordevic2021}. The need to cool and trap the atoms imposes a complex optical infrastructure, bulky ultra-high vacuum systems, and significant deadtime making the data rate quite low.

These complications have sparked renewed interest in examining room-temperature atomic vapors or fast atomic beams as a means of atomic-cavity-QED. Using hot atoms has shown the ability to provide technological and scientific solutions, including phase coherent atomic driving \cite{Kim2018}, mapping the vacuum state of the cavity mode \cite{Lee2014}, and performing optical magnetometry \cite{Budker2007}. To enhance scalability, many groups are pursuing integrated chip-scale devices that interact with atomic vapors \cite{Kitching2018, Zektzer2024, Stern2013, Cutler2020} with applications such as low-power optical switching \cite{Stern2016}, small length scale magnetometry \cite{Gerginov2020}, and compact optical spectrometers\cite{Edrei2022, Hummon:18}.

Of these chip-scale devices, there has been significant progress in realizing scalable fabrication of small atomic vapor cells \cite{Liew2004} for electrometry \cite{Sedlacek2012}, inertial sensing \cite{Walker2016}, and atomic clocks \cite{Knappe2004, Lutwak2007}. A recent innovation has been the development of a fully enclosed and compact Rb beam source, whose collimation is achieved using etched micron-sized channels \cite{Li2019, Li2020}. This type of source was used recently to realize a chip-scale microwave atomic beam clock using Ramsey coherent population trapping \cite{Martinez2023}.




Here, we demonstrate that such a beam source is compatible with performing cavity-QED with a high finesse optical cavity. We demonstrate that the source does not degrade the cavity and that we can realize strong collective light-atom coupling, few-photon optical non-linearity, and the generation of non-classical light. Our approach is a critical step in enabling the future integration of optical cavities into compact and low-volume atomic cavity-QED systems with vastly reduced complexity in size, weight, power, and cost when compared to cold atom cavity-QED systems. Further, this work demonstrates the future feasibility of incorporating medium to high finesse cavity mirrors into miniature microwave clocks and matterwave interferometers for enhanced signal-to-noise.   



\section*{Experimental setup}


\begin{figure*}[!ht]
	\centering
    \includegraphics[width=\textwidth]{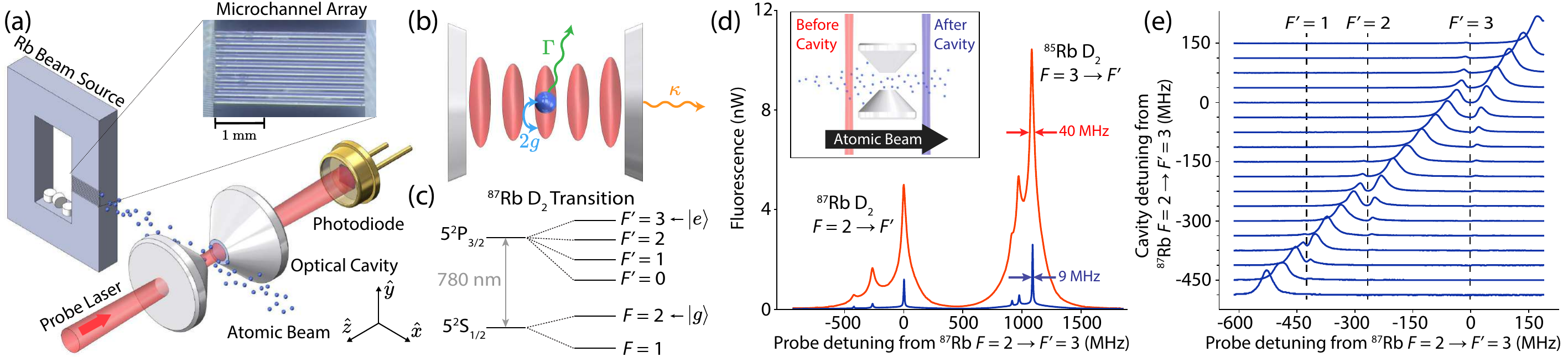}
	\caption{\textbf{Experimental Setup.} \textbf{(a)} The compact atomic source (gray box) produces a thermal atomic beam (blue spheres) by feeding an atomic vapor of Rb through a microchannel array (inset), which propagates through an optical cavity formed by two high-finesse mirrors (white cones). Atoms are excited using a weak probe beam, while the photons transmitted through the cavity are collected using a photodiode (gold). \textbf{(b)} Diagram of the atom-cavity interactions and dissipation mechanisms. $^{87}$Rb atoms (blue spheres) in $F=2$ are coupled to the cavity mode (red ovals) at a vacuum Rabi frequency of $2g$. Excitations in the atom-cavity system are dissipated through atomic spontaneous emission at rate $\Gamma$ or cavity power decay at rate $\kappa$. \textbf{(c)} $^{87}$Rb D$_{2}$ transition hyperfine structure. \textbf{(d)} Fluorescence spectroscopy of the atomic beam perpendicular to its direction of propagation. The $^{87}$Rb and $^{85}$Rb D$_2$ transitions are characterized $\qty{10}{\mm}$ before (red) and after (blue) the optical cavity. \textbf{(e)} The normal mode splitting of the atom-cavity system on the three allowed $^{87}$Rb $F = 2 \rightarrow F'$ hyperfine transitions.}
	\label{fig1}
\end{figure*}

\subsection*{Atomic source}

The miniature atomic beam source is shown in Fig. ~\ref{fig1}a. The device is a modification of the chip-scale atomic beam devices which have been used to realize a miniature atomic beam clock \cite{Li2019,Martinez2023}, here adapted to provide an atomic beam through the center of an optical cavity. The beam source consists of a silicon wafer with an internal reservoir that contains a Rb vapor. A microchannel array is etched into the silicon surface and is fed by the vapor to produce a series of atomic beams along the axis of the microchannels. The device is enclosed by two layers of borosilicate glass. One glass layer is anodically bonded to the silicon to form a hermetic seal, while the other is mechanically affixed to the structure to allow for easy loading and reloading of alkali metal into the device. The beam source is placed inside an actively pumped vacuum chamber, and the atomic beams are aligned with the cavity opening. 


The microchannel array consists of 10 rectangular channels spaced by $\qty{150}{\um}$ center to center. Each channel has a length of $\qty{3}{\mm}$ and a square cross-section of $\qty{100}{\um}$ by $\qty{100}{\um}$. The channels' aspect ratio (ratio of length to width) is 30, which creates atomic beams with a half-angle divergence of roughly 33 milliradans (mrad) \cite{Beijerinck1975}. Rubidium is introduced into the beam device using Rb molybdate Zr/Al pill-type dispensers which are thermally activated using laser heating to produce pure Rb metal~\cite{douahiVapourMicrocellChip2007a} (See Methods).  Once the Rb is deposited, the beam device is resistively heated up to $180^\circ$~C to control the Rb vapor pressure within the vapor cell region. For most of the work here, we operate at source temperatures between 40 and $70^\circ$ C, where the total atomic flux from the array varies from $10^{10} \ \text{s}^{-1}$ to $2\times10^{11} \ \text{s}^{-1}$. Each Rb pill contains 0.4 mg of natural abundance Rb, and this is expected to sustain source operation at $70^\circ$ C for more than 5 months. We observe stable operation of the beam source over roughly 3 months with a single activation of the pills. 


\subsection*{Cavity design}



Our Fabry–Pérot microcavity (See Methods) is similar to those of  \cite{ Mabuchi1999, Sauer2004, Khudaverdyan2009, Mucke2010}. More advanced and integrated designs \cite{Hunger2010, Uphoff2016, McLemore2024, grinkemeyer2024, Bao:17} could be considered for future work, but the goal in the present context is to understand if low-transmission mirror coatings that are common to all of these other approaches are compatible with the miniature beam source at a level sufficient to produce nonclassical light fields.

For the results presented here, we used two different cavities that we will refer to as A and B. All mirrors used were either low or high transmission, with power transmission of $2\times 10^{-6}$ and $40\times 10^{-6}$ respectively at the atomic transition wavelength $\lambda_a \approx 780.2$~nm. For cavity A, we use two low transmission mirrors so that light exits the cavity equally from both mirrors. For cavity B, we use one low and one high transmission mirror so that most of the light exits via the high transmission end. Both cavity lengths were set to $L= \qty{80(5)}{\um}$, and their resonance frequencies were tuned near the $ ^{87}$Rb $D_2$ $\ket{5^2 S_{1/2}, F = 2} \rightarrow \ket{5^2 P_{3/2}, F' = 3,2,1}$ transitions (Fig.~\ref{fig1}c), which have a decay rate of $\Gamma \approx 2\pi \times \qty{6.1}{\MHz}$ \cite{Volz1996}. The TEM$_{00}$ mode had a waist of $w_0 = \qty{22.3(4)}{\um}$ at $\lambda_a$. For cavities A and B, respectively, the measured cavity linewidths are $\kappa/2 \pi = 33(1)$ and $13(1)$~MHz, indicating per mirror power loss coefficients of $ 15(3) \times 10^{-6}$ and $ 1(1) \times 10^{-6}$. Their corresponding finesses are $\mathcal{F}_A = 0.6(1) \times 10^5$ and $\mathcal{F}_B = 1.5(1) \times 10^5$.

The atom-cavity system operates in the strong-coupling limit as characterized by the single-particle cooperativity $C = 4g_0^2/\kappa\Gamma \gg 1$, where $2g_0$ is the vacuum Rabi frequency describing the coherent atom-cavity interactions \cite{Kimble1998}. In the limit $\kappa\gg 2g_0,\Gamma$, the cooperativity characterizes the probability $P_c = C/(1+C)$ that an atom in the optically excited state decays to the ground state by emitting a photon into the cavity \cite{Tanji2011}. Given the cavity effective mode volume $V=w_0^2 L/4$ and dipole moment for the $ ^{87}$Rb $D_2$ cycling transition $\ket{F = 2, m_{F} = 2} \rightarrow \ket{F' = 3, m_{F'} = 3}$, the vacuum Rabi frequency at an antinode on the cavity axis is $2 g_0 = 2\pi \times \qty{52(2)}{\MHz}$, yielding peak cooperativities of $C=18(2)$ and $35(3)$ for cavities A and B, respectively. As the atoms transit the standing wave cavity mode, the coupling will vary with position as

\begin{equation}
    g(\vec{r}) = g_0\cos\left(\frac{2\pi z}{\lambda_{a}}\right)e^{-\frac{x^2+y^2}{w_0^2}},
\end{equation}

\noindent where $z$ is the position along the cavity axis and $x$ is the position along the atomic beam direction. To account for the inhomogeneous coupling of atoms within the standing wave profile of the cavity, we define an effective vacuum Rabi frequency $2g_{\mathrm{eff}} = \sqrt{3/8} \times 2 g_0 = 2\pi \times \qty{32(2)}{\MHz}$ and a time-averaged effective atom number $N_{\mathrm{eff}}$ for a two-level atom (See Methods), in a fashion reminiscent to the effective couplings used in cavity-aided spin squeezing \cite{Zilong2011, Leroux2010, Schleier2010}.

\begin{figure}[t]
	\centering
	\includegraphics[width=\columnwidth]{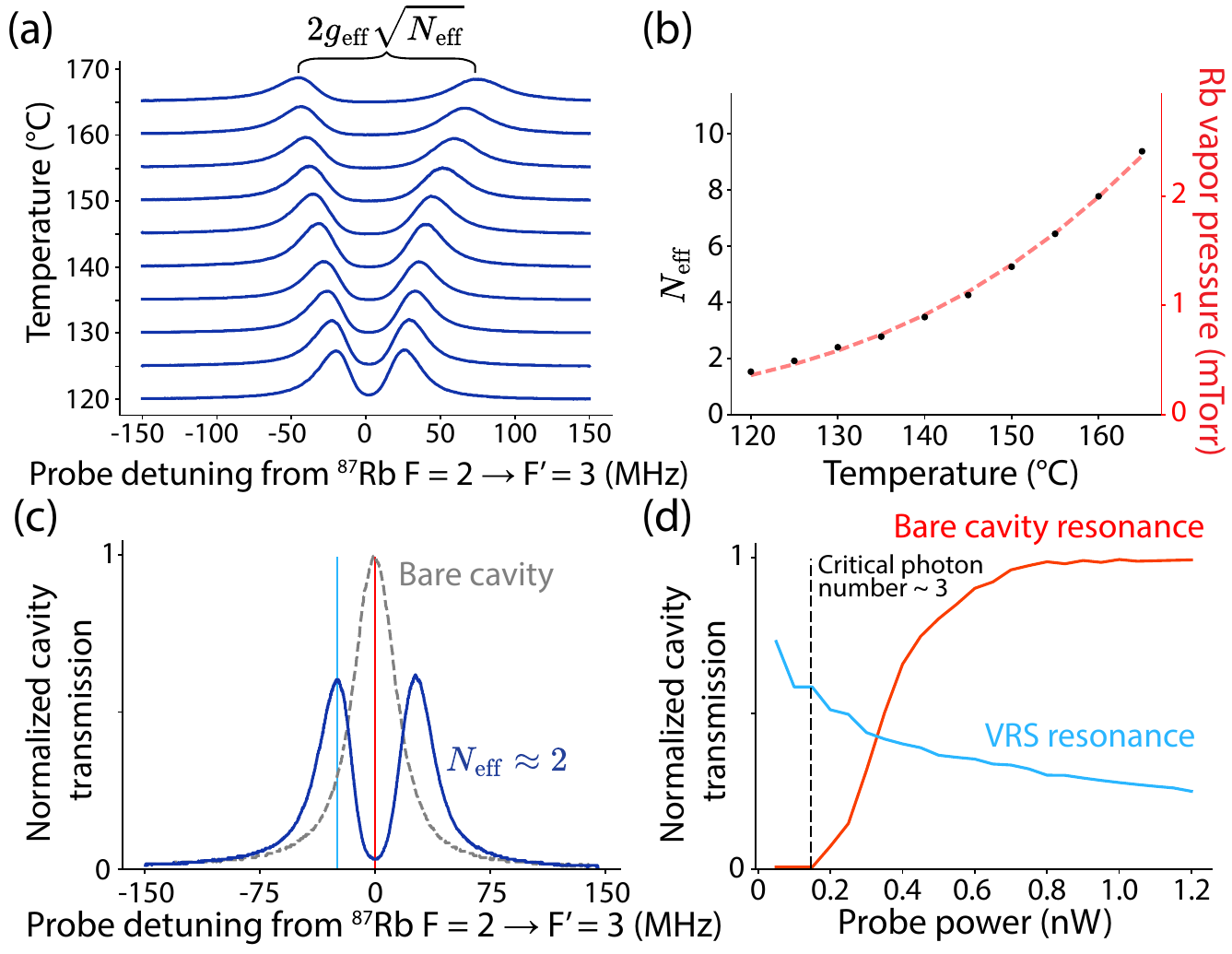}
	\caption{\textbf{Demonstration of the cavity QED platform.} \textbf{(a)} Splitting of the VRS peaks increases as a function of the source temperature. \textbf{(b)} The temperature dependence of $N_{\mathrm{eff}}$ (black dots) agrees with the Rb vapor pressure curve (red dashed line) up to a multiplicative scaling factor. \textbf{(c)} The transmission in (c) and (d) is normalized to the peak transmission of the bare cavity. Bare cavity transmission as a function of probe detuning is shown in grey. Transmission with $N_{\mathrm{eff}} \approx 2$ is shown in blue, showing VRS peaks. \textbf{(d)} Nonlinear cavity transmission as a function of the probe power. The blue and red curves are taken at detunings corresponding to the blue and red lines in (c). This data was also taken with $N_{\mathrm{eff}} \approx 2$.}
	\label{fig2}
\end{figure}

\subsection*{Interfacing the source with the cavity}


The output of the Rb source is $\qty{10}{\mm}$ from the cavity mode. The angular alignment of the atomic beam is precise to within a few mrad, tuned using a dispersive feature in the cavity probe transmission (see Methods and \cite{Famà2024}).
The mean thermal velocity of $v \approx \qty{290}{\m/s}$ at $70^\circ$~C gives an average cavity transit time of $t_{T} \approx \qty{150}{\ns}$ across the cavity mode diameter $2 w_0$. The root mean square (rms) angular divergence of the atomic beam is calculated from the linewidth of the $ ^{85}$Rb $F = 3 \rightarrow F' = 4$ transition as measured using fluorescence spectroscopy (Fig.~\ref{fig1}d and Methods). Just before the cavity, the rms angular divergence of the full atomic beam is $\qty{43(3)}{\milli\radian}$. However, the portion of the beam that passes through the cavity has an rms angular divergence of only $\qty{8(2)}{\milli\radian}$ such that the rms displacement along the cavity axis $\hat{z}$ is only $0.23 \lambda_a$ per mode waist $w_0$ traveled along $\hat{x}$. 





\subsection*{Atom-cavity coupling}

As seen in Fig.~\ref{fig1}e, the atom-cavity coupling hybridizes the atomic and cavity modes, leading to a series of avoided level crossings in the measured cavity-probe transmission of Cavity A (Fig.~\ref{fig1}a) versus probe frequency, with the cavity resonance frequency $\omega_c$ varied over a range spanning the $ ^{87}$Rb $\ket{F = 2} \rightarrow \ket{F' = 3,2,1}$ transitions between sweeps. As the cavity comes into resonance with each transition, the single cavity resonance frequency splits into two new normal modes. 

In Fig.~\ref{fig2}a, we observe the hybridized modes (sometimes called polaritons) when we set the cavity to resonance with the $\ket{F = 2} \rightarrow \ket{F' = 3}$  transition $\omega_a$. $N_{\mathrm{eff}}$ and $g_{\mathrm{eff}}$ are defined so that the frequency splitting or vacuum Rabi splitting (VRS) of the two hybridized modes is $2g_{\mathrm{eff}} \sqrt{N_{\mathrm{eff}}}$. We vary the Rb source temperature and measure the splitting to determine $N_{\mathrm{eff}}$ as a function of temperature. The scaling of $N_{\mathrm{eff}}$ matches the behavior of the known Rb vapor pressure curve \cite{Nesmeianov1963} up to a constant scale factor (Fig.~\ref{fig2}b).  We can also resolve the separation of the VRS peaks for $N_{\mathrm{eff}} \gtrsim 1$. The full width at half maximum (FWHM) linewidth of each hybridized normal mode is predicted to be $(\kappa + \Gamma)/2 = 2\pi \times \qty{19.5(5)}{\MHz}$, compared to the measured linewidth $\qty{22(1)}{\MHz}$, which we attribute to additional broadening from Poisson variance in the atom number. 


Unlike the cavity mode, each atom can carry only one optical excitation.  This is reflected in the nonlinear transmitted power versus probe power beginning at an intracavity photon number of roughly 3(1) (Fig.~\ref{fig2}c, Fig.~\ref{fig2}d), known as the critical photon number $n_0$ \cite{Kimble1998, Gripp1996, GeaBanacloche2008}. This few-photon optical non-linearity could be used for optical switching at the few photon level.

\section*{Non-classical light}

\subsection*{Second-order correlation measurement}

\begin{figure}[t]
	\centering
	\includegraphics[width=\columnwidth]{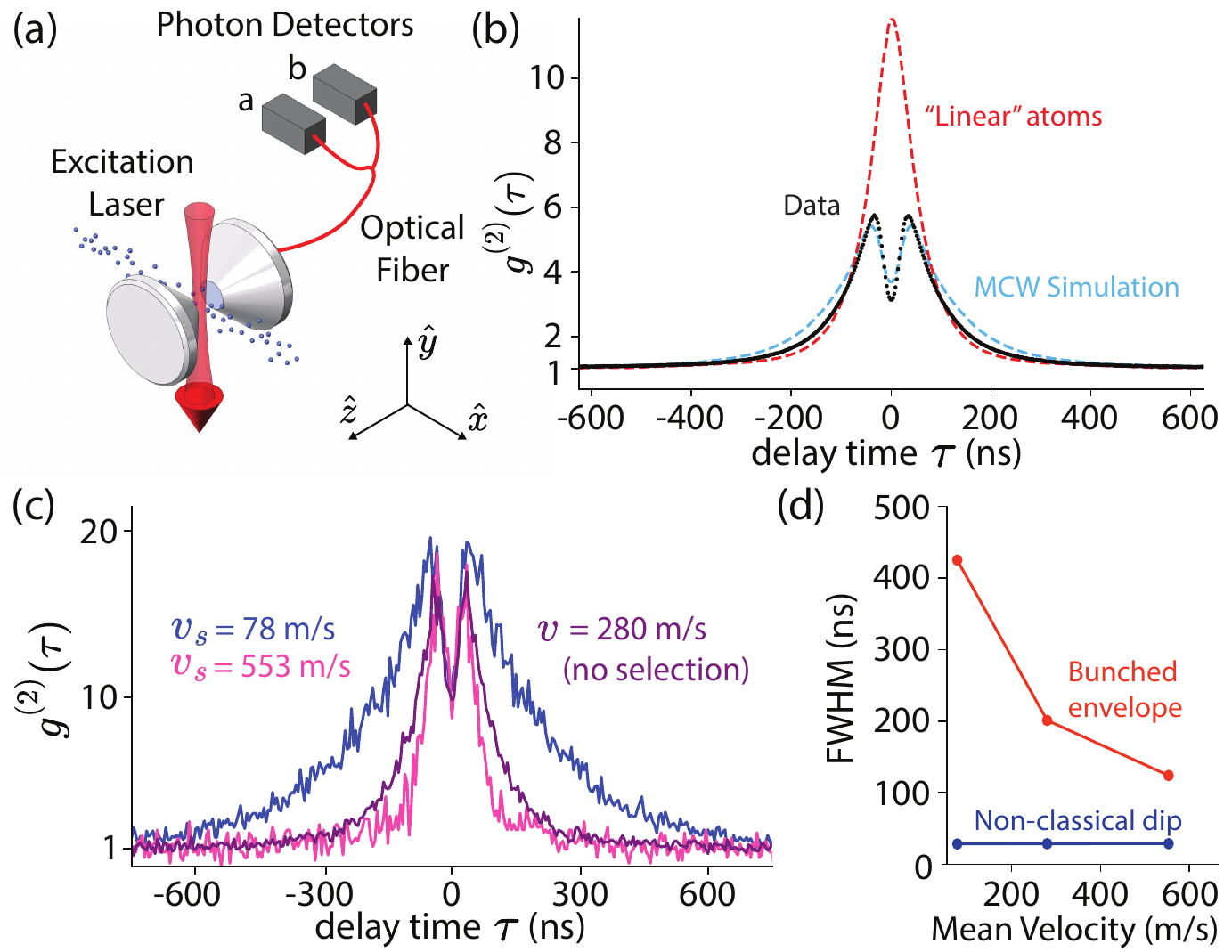}
	\caption{\textbf{Non-classical light.} \textbf{(a)} An excitation beam overlaps with the cavity mode to continuously excite atoms during their transits. The cavity output is collected on two SPCMs. \textbf{(b)} Measured profile of $g^{(2)}(\tau)$ (black dots). The $g^{(2)}(\tau)$ displays a bunched profile as well as a non-classical dip within the bunched envelope that is centered at $\tau = 0$. The dashed blue line is  a Monte Carlo Wavefunction (MCW) simulation, the details of which are explained in the Methods. The dashed red line is a classical model for the bunched envelope, ignoring the nonlinearity of the atoms. \textbf{(c)} Velocity-selective pumping. Relying on the Doppler shift, we use optical pumping to pump atoms into F = 2 with a desired velocity. Profiles of $g^{(2)}(\tau)$ are shown with no optical pumping (purple), $v_s \approx 78~\mathrm{m/s}$ (blue), and $v_s \approx 553~\mathrm{m/s}$ (pink). \textbf{(d)} The width of the bunched envelope changes to reflect the transit time, while the width of the non-classical dip is constant to within measurement uncertainty.
    }
	\label{fig3}
\end{figure}

The nonlinearity of the atoms is next used to realize a non-classical light source. We reduce the Rb source temperature to lower the effective atom number to $N_{\mathrm{eff}} \approx 0.1$, limiting the likelihood of two atoms transiting the cavity mode at the same time.  Cavity B's resonance frequency is tuned to resonance with $\omega_a$.  The atoms are now directly excited at frequency $\omega_a$ using a laser beam along $-\hat{y}$ (see Fig.~\ref{fig3}a) with a Rabi frequency of $\Omega = 2 \pi \times 36^{+ 5}_{-18}~\mathrm{MHz}$. 

Photons exiting the cavity are sent to a nonpolarizing 50/50 beamsplitter, and each output is directed to a single photon counting module (SPCM). The detected photons' statistics are characterized using the second-order correlation function $g^{(2)}(\tau)$ \cite{Glauber1963}. Given the detection of a photon at time $t$, $g^{(2)}(\tau)$ gives the relative likelihood to detect a second photon at time $t+\tau$ compared to what we would expect if the arrival times of photons were uncorrelated, or Poissonian. $g^{(2)}(\tau) > 1$ indicates that a second photon is \textit{more} likely to arrive than for Poissonian light, whereas $g^{(2)}(\tau) < 1$ indicates that a second photon is \textit{less} likely to arrive than for Poissonian light.

The measured $g^{(2)}(\tau)$ in Fig.~\ref{fig3}b features a bunched envelope that asymptotes to 1 at large $\tau$.  Within this envelope, a central feature appears as a dip with a FWHM of $t_{\mathrm{dip}} = \qty{28(3)}{\mathrm{ns}}$ and a local minimum $g^{(2)}(0) =  3.13(1)$. This feature clearly violates the classical inequality $g^{(2)}(0) > g^{(2)}(\tau)$, certifying that the light source is non-classical \cite{Foster2000}.  The dip conceptually arises from the time scale for the atom to return to the optically excited state after it has emitted a photon. The minimum value of $g^{(2)}(0)$ is limited by the probability of multiple photons being deposited in the cavity at the same time. This can happen due to either the finite storage time $1/\kappa$ of the cavity or due to multiple atoms transiting the cavity at once.

The characteristic timescale for the bunching envelope is that of the atom number fluctuations, or the average atom transit time $t_T\approx \qty{150}{\ns}$ \cite{Foster2000}.  The red dashed curve of Fig.~\ref{fig3}b shows the expected $g^{(2)}(\tau)$ if the atoms were classical linear scatterers transiting the cavity with random atom arrival times, following the thermal atomic velocity distribution and the measured angular beam divergence along the cavity axis. This model generally reproduces the shape of the bunched envelope, but does not display a central dip.

Modeling the full shape requires capturing the nonlinearity of the atoms, which we do using a Monte Carlo wave function simulation (the dashed blue line in Fig.~\ref{fig3}b) (see Methods). The values of $g_0$ and $\Omega$ are adjusted to better describe the observed dip. We also observe how varying the cavity linewidth and atomic beam flux affects $g^{(2)}(\tau)$. These simulations support the physical intuition that the finite cavity-linewidth $\kappa$ is the primary limitation on the minimum value of $g^{(2)}(0)$ because the associated storage time $1/\kappa$ blurs the time dynamics.



We can vary the interaction time with the cavity using velocity-selective optical pumping of atoms from the ground state $F=1$ (which does not resonantly interact with the cavity) to $F=2$ (which does resonantly interact with the cavity). In Figure \ref{fig3}c, we show the measured $g^{(2)}(\tau)$ with no selection and for $v_s \approx 78$ and 553~m/s. The observed bunching timescale changes by $\approx 2$ without changing the width of the non-classical central feature, as shown in Fig.~\ref{fig3}d.


\begin{figure*}[t]
	\centering
	\includegraphics[width=\textwidth]{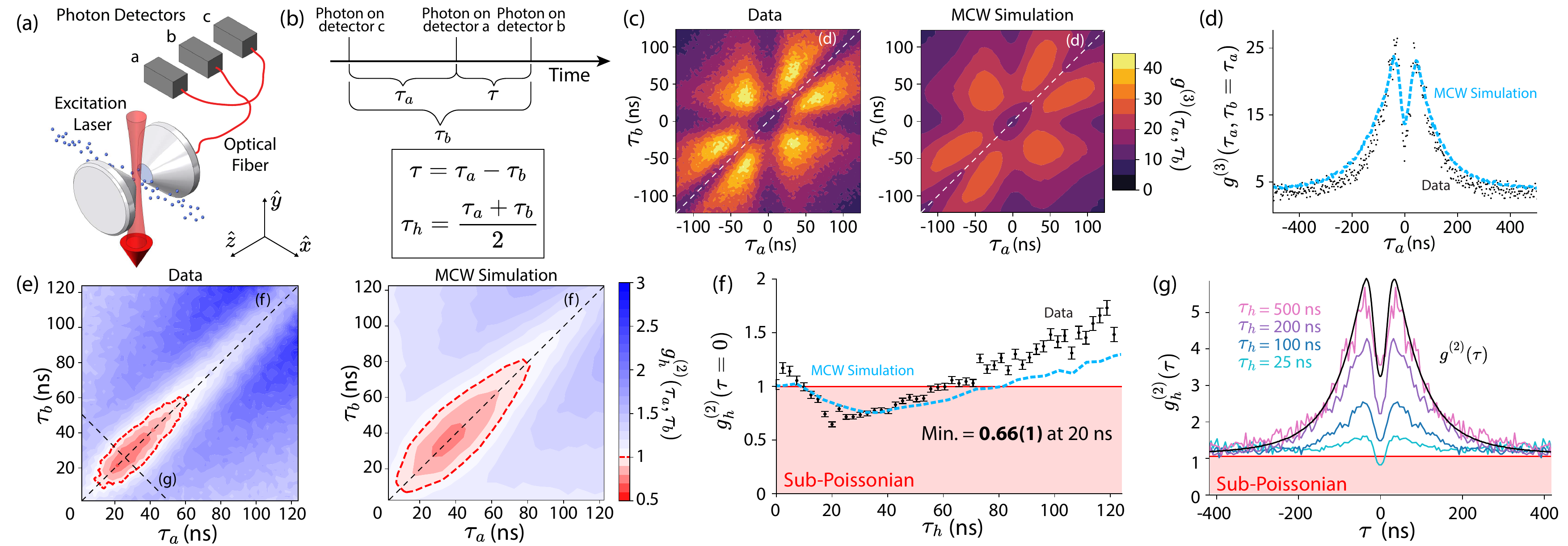}
	\caption{\textbf{$g^{(3)}(\tau_a, \tau_b)$ and heralded $g^{(2)}(\tau)$ measurement.} 
    \textbf{(a)} To measure $g^{(3)}(\tau_a, \tau_b)$ and heralded $g^{(2)}(\tau)$, we add a third SPCM.
    \textbf{(b)} To compare $g^{(2)}_h(\tau_a, \tau_b)$ to non-heralded $g^{(2)}(\tau)$, we redefine our variables: $\tau_h$ is the wait time after the herald photon, and $\tau$ is the time difference between the two heralded photons.
    \textbf{(c)} Three-photon correlation $g^{(3)}(\tau_a, \tau_b)$: data (left) and theory (right). It can be seen that $g^{(3)}(0,0) < g^{(3)}(\tau_a, \tau_b)$ for some nonzero values of $\tau_a, \tau_b$, which is also an indicator of non-classical light.
    \textbf{(d)} A profile along the diagonal of (c) with $\tau_a = \tau_b$ again shows $g^{(3)}(0,0) < g^{(3)}(\tau_a, \tau_b)$. The blue dashed curve shows the same profile for the theory.
    \textbf{(e)} A full two-dimensional plot of $g^{(2)}_h(\tau_a, \tau_b)$: data (left) and theory (right). The red region along the diagonal has $g^{(2)}_h(\tau_a, \tau_b) < 1$. 
    \textbf{(f)} A profile along the diagonal of (e) can be interpreted as showing $g^{(2)}(0)$ as a function of the wait time $\tau_h$ after the herald photon. $g^{(2)}(0)$ drops below 1 for roughly $\tau_r < \tau_h < t_T$. The blue dashed curve shows the same profile for the theory. \textbf{(g)} As an alternate perspective, profiles of $g^{(2)}_h(\tau)$ along lines of constant $\tau_h$ drop below 1 for small $\tau_h$ but approach the non-heralded $g^{(2)}(\tau)$ in the limit of $\tau_h \gg t_T$. In this limit, the post-selection is on random time windows rather than specifically times when there is an atom in the cavity, so we recover the non-heralded bunching behavior. 
    }
	\label{fig4}
\end{figure*}

\subsection*{Third-order correlation measurement}

The thermal atomic beam source affords a high data rate of $\approx$10$^5$ photon counts per second at 65$^\circ$C, which is integrated for 8 hours to produce the data shown in Figs.~\ref{fig3} and~\ref{fig4}. This is sufficient to characterize the photons' third-order correlation, $g^{(3)}(\tau_a, \tau_b)$ \cite{Glauber1963} (see Fig.~\ref{fig4}(a, b)), using three SPCMs labeled {a,b,c} to avoid detector dead time. The measured $g^{(3)}(\tau_a, \tau_b)$ shown in Fig.~\ref{fig4}c also displays bunching due to the random transit times of the atoms through the cavity. The six-fold symmetry corresponds to the six permutations of labels for the three detected photons. Classical light obeys the relation $g^{(3)}(\tau_a, \tau_b) \le g^{(3)}(0, 0) $, which our measured $g^{(3)}(\tau_a, \tau_b)$ violates as seen in Fig.~\ref{fig4}c, and as emphasized by a cut along the white dashed line $\tau_a=\tau_b$ shown in Fig.~\ref{fig4}d. For comparison, the results in Fig.~\ref{fig4} are in good agreement with the corresponding correlation functions computed using the same MCWF simulations from the previous section.

\subsection*{Heralded second-order correlation}


With a thermal atomic beam, we do not have the degree of control over the atoms that would be needed to suppress fluctuations in atom number. However, because we are not probing along the cavity axis, the detection of a photon from the cavity acts as a signal that there is an atom in the cavity. Thus, we can post-select on the presence of an atom in the cavity by requiring the detection of an initial cavity photon. In practice, this allows us to use the $g^{(3)}(\tau_a, \tau_b)$ data to form a heralded second-order correlation measurement \cite{Signorini2020, Razavi2009} shown in Fig.~\ref{fig4}e that is sub-Poissonian $g^{(2)}_h(\tau_a, \tau_b)<1$ within the dashed red region. Explicitly, the heralded correlation function is calculated as

\begin{equation}
g^{(2)}_h(\tau_a, \tau_b) = \frac{M_{cab}(\tau_a, \tau_b)M_c}{M_{ca}(\tau_a)M_{cb}(\tau_b)}.
\end{equation}

Here detector $c$ is taken to be herald detector, and correlations are calculated between the other two detectors $a$ and $b$ conditioned on the prior arrival of a photon in $c$. The heralded second-order correlation is a function of two times, $\tau_a$ being the time difference between photons $c$ and $a$, and $\tau_b$ between photons $c$ and $b$ (Fig.~\ref{fig4}b). Each $M$ represents the number of single ($M_c$), double ($M_{ca},M_{cb}$) or triple photon detections ($M_{cab}$), with double or triple coincidences occurring on different detectors at time differences $\tau_a$ and $\tau_b$ (see Methods).




The diagonal cut shown in Fig.~\ref{fig4}f is equivalent to a heralded second order correlation function at $\tau\equiv\tau_a - \tau_b =0$.  The minimum observed $g_h^{(2)}(0)$ is 0.66(1) at $\tau_h = \qty{20}{\ns}$, confirming that we have sub-Poissonian light.  The minimum occurs at $\tau_h \neq 0$ since the atom requires a finite reset time after having emitted the herald photon.  At larger times $\tau_h$, the atom has left the cavity such that the heralding no longer applies and one returns to a bunching signal once again.

The opposite diagonal cuts shown in Fig.~\ref{fig4}g at fixed herald times $\tau_h$ is equivalent to a $g_h^{(2)}(\tau)$.  One sees that the light is antibunched at $\tau=0$ for $\tau_h=\qty{25}{\ns}$ with reduced bunching away from $\tau=0$.  The antibunching disappears at larger  heralding times $\tau_h$, returning to the previous unheralded result (black) at $\tau_h = 500~\mathrm{ns}\gg t_T$.


\section*{Outlook}



These results demonstrate that the miniature and mass-producible rubidium beam source is compatible with achieving $C\gg1$ without degradation of the optical cavity over time due to Rb adhering to the mirror surface. 
With the source active between 40 and $70^\circ$ C for 6 months, we have not observed a discernible change in the finesse of the cavity at the level of $\qty{10}{}\%$. This implies a maximum increase in losses of $0.5 \times 10^{-6}$ per mirror per month, which in a pessimistic linear model is sufficient to operate a cavity with a finesse of $10^5$ for three years without being loss dominated.

Ultimately, this experiment serves as a first step towards creating a miniaturized cavity QED system with a thermal atomic beam. While our atomic source is chip-scale, our optical cavity is freestanding and enclosed within a larger vacuum chamber. In the future, we will explore the integration of these cavities into sealed and passively vacuum pumped chip scale devices such as in previously successful work \cite{Martinez2023}.  Such a device will find broad applications in miniature atomic clocks, chip-scale magnetometers, low-power optical switching, and generation of non-classical light. 

\bibliography{NonclassicalLight}

\section*{Methods}

\subsection{Cavity Details}


The concave mirrors with curvature $R_c=10$~cm and diameter of $D_m=7.75$~mm geometrically limit the on-axis mirror separation $L$ to be $L\ge D_m^2/4R_c = \qty{150}{\um}$ so that the mirrors do not touch at their edges, leaving no gap for atoms to enter the cavity. To achieve smaller mirror separations, the 4~mm thick fused silica substrates were machined to form a cone starting from their base diameter of 7.75~mm down to a diameter of $D_m=\qty{1.5}{\mm}$ at the mirror surface. In principle, this allows for a minimum cavity length of $L = \qty{6}{\um}$ at zero gap.

A CNC machine is used to cone down each mirror to the desired diameter. First, the AR-coated side of the mirror is waxed onto a positioning fixture with Pelco Quickstick$^{\mathrm{TM}}$ 135 Mounting Wax. The high-finesse surface is then sprayed with several layers of clear acrylic (Kryon Industrial 
Acryli-Quik$^{\mathrm{TM}}$) to protect it during machining. Starting $\qty{0.5}{mm}$ from the AR-coated surface, a medium grit diamond grinding wheel is used to cut the substrate at an angle of $41.8^{\circ}$ from the normal of the mirror surface. 

We perform twelve different cuts in four sequences to bring the substrate down to a cone with a diameter of $\qty{1.5}{\mm}$ at the high-finesse surface. We check that the mirror is still waxed on the fixture between each cut. After machining, acetone dissolves the wax and acrylic from their respective surfaces.




\subsection{Depositing Rubidium in Source Region}
Rubidium is deposited in the source region using a 4 W, 975 nm laser from QPhotonics to heat AlkaMax pills from SAES for 15 minutes. After the the pills are activated, we use a resistive cartridge heater (Thorlabs HT15W) to heat the entire device to temperatures between $40^\circ$~C and $180^\circ$~C. This heating adjusts the vapor pressure of the rubidium trapped within the source, which determines the flux of rubidium leaving through the micro-channel array. The mean thermal velocity of our atomic beam flux varies from $\approx$ $\qty{275}{\m/s}$ to $\qty{330}{\m/s}$ over our operating temperature range. The pills can be activated up to 10 times using this procedure before they are depleted.



\subsection{Divergence of Atomic Beam}

To characterize the angular divergence of the atomic beam before and after passing through the cavity, we perform fluorescence spectroscopy \qty{10}{mm} before and after the optical cavity. For the $ ^{85}$Rb $F = 3 \rightarrow F' = 4$ transition before the cavity, the linewidth is measured to be $\qty{40}{\MHz}$ in agreement with \cite{Martinez2023}. This is equivalent to an rms angular divergence of $\qty{43(3)}{\milli\radian}$. A similar measurement on the atoms exiting the cavity gives a much smaller linewidth of $\qty{9}{\MHz}$ and an rms angular divergence of $\qty{8(2)}{\milli\radian}$. The narrow cavity acts as a spatial filter, selecting a population of atoms with less angular divergence to interact with the cavity mode.

\subsection{Atomic Beam Alignment}

To ensure that the angular distribution of the atomic beam is centered with respect to the optical cavity mode, we monitor an additional dispersive feature that appears in the probe transmission when the atomic beam passes through the cavity at an angle \cite{Famà2024}. The dispersive feature arises from a modification of the round trip phase of the cavity's circulating field by atoms that experience a Doppler shift of the cavity field. This phase shift produces a third resonant condition for the cavity, which is only partially attenuated by absorption from the atomic ensemble. 

To align the center of the atomic beam with the optical cavity mode, we use a Thorlabs PDR1V translation stage to move the source along the optical cavity axis. When the atomic beam is centered on the cavity axis, this dispersive feature disappears as the cavity field experiences two equal but opposite phase shifts from atoms traveling with equal velocities in opposite directions.

\subsection{Cavity Stabilization}
The cavity frequency is stabilized using a Pound-Drever-Hall (PDH) lock to a laser that is in turn locked at variable frequency offset from the $^{85}$Rb D2 $F=3$ to $F=4$ transition. The PDH laser light is sufficiently strong to saturate any atoms in the cavity, allowing us to measure the bare cavity-resonance frequency. 

The PDH laser light is alternately toggled on and off for $\qty{220}{\us}$, faster than the unity gain frequency of the cavity feedback loop.  Science is done during the off period, during which the field emitted from the opposite end of the cavity is detected on 2 to 3 SPCMs. The SPCMs are gated off during the PDH probe period.

\subsection{Velocity-Selective Optical Pumping}

To verify that the characteristic timescale of the bunching corresponds to the average atom transit time, we use velocity-selective optical pumping to modify the shape of the peak. After optically pumping the population from $\ket{F = 2}$ to $\ket{F = 1}$, a beam angled $\theta = 58(1)^{\circ}$ from the horizontal with detuning $\delta$ from the $\ket{F = 1} \rightarrow \ket{F' = 2}$ transition is used to pump the atoms back into $\ket{F = 2}$. 

Since the velocity-selective pump has a component parallel to the atomic beam, we can select atoms moving at velocity $v_s$ by choosing $\delta =v_s\cos(\theta)/\qty{780}{nm}$ to the red of the transition. The width of the velocity selection is broadened by excitations of the atomic beam by the vertical component of the optical drive as well as through velocity classes that are excited on the $\ket{F = 2} \rightarrow \ket{F' = 1}$ transition.

\subsection{Theoretical Model and Monte-Carlo Wavefunction Simulation}

The Hamiltonian describing a total of $N_t$ atoms transiting the cavity will be approximated by a two-level atomic Hamiltonian with transition frequency $\omega_a$, and an excitation laser at frequency $\omega_d$ and Rabi frequency $\Omega_i(t)$ as

\begin{equation}
    \hat{H}_a = \frac{1}{2}\sum_{i=1}^{N_t} \left(\omega_z \hat{\sigma}^z_i+\frac{1}{2}\left(\Omega_i(t)\hat{\sigma}_i^+ + \Omega_i^{*}(t)\hat{\sigma}_i^-\right)\right)
\end{equation}

\noindent where $\hat{\sigma}_i^\alpha$ is the usual Pauli spin operator for the $i^{th}$ atom with $\alpha= \left\{x, y, z, +, -\right\}$.

The cavity is described by a quantum harmonic oscillator Hamiltonian 

\begin{equation}
    \hat{H}_c = \omega_c \left(\hat{c}^\dagger \hat{c} + \frac{1}{2}\right),
\end{equation}

\noindent where $\hat{c}$ and $\hat{c}^\dagger$ are bosonic creation and annihilation operators for the cavity field.

For non-classical light generation the cavity and drive are both tuned to resonance with the atomic transition frequency so that $\omega_c=\omega_d=\omega_a$. Moving into a rotating frame at $\omega_a$, and including the Tavis-Cummings atom-cavity interaction \cite{Tavis1968}, we have a total Hamiltonian

\begin{equation}
    \hat{H}_t = \sum_{i=1}^{N_t} \left[g_i\left(t\right)\left(\hat{\sigma}_i^+ \hat{c} + \hat{\sigma}_i^- \hat{c}^\dagger \right) + \frac{1}{2}\left(\Omega_i(t) \hat{\sigma}_{i}^{+} + \Omega_i^{*}(t)\hat{\sigma}_i^-\right) \right]
\end{equation}

\noindent where the time-varying components are

\begin{equation}
    g_i(t) = g_0 e^{-\frac{x_i^2(t)+y_i^2(t)}{w_0^2}}\cos\left(\frac{2 \pi z_i(t)}{\lambda_a}\right)
\end{equation}
and 
\begin{equation}
\Omega_i \left(t\right) = \Omega e^{2\pi \imath y_i(t)/\lambda_a}
\end{equation} 

\noindent which represent the coupling of the $i^{th}$ atom to the standing wave cavity mode and excitation laser, respectively. The atoms are approximated as following classical ballistic trajectories $\vec{r}_i(t) = x_i(t)\hat{x}+y_i(t)\hat{y}+z_i(t)\hat{z}$. Cavity decay at rate $\kappa$ and single-atom decay $\Gamma$ from the excited state via spontaneous emission into free space are described by jump operators  $\sqrt{\kappa}c$ and $\sqrt{\Gamma}\sigma_i^-$, respectively. 






Monte-Carlo Wavefunction (MCWF) simulations are performed using QuTiP \cite{lambert2024} and compared to data in Fig.~3 and 4 of the main text. We use a truncated Hilbert space that allows for a maximum of two atoms and five photons in the cavity at once. The parameters of the system are such that it is very rare for these limits to be exceeded. 

The simulated atomic beam flux is chosen to match the amplitude of the bunched envelope, and the parameters $g_0= 2 \pi \times 15$~MHz and $\Omega = 2\pi \times 7$~MHz for a two-level atom are adjusted in the simulation to more accurately describe the central dip. These values should be compared with the predicted values for a multi-level atom $g_0 = 2 \pi \times 22(2)~\mathrm{MHz}$ and $\Omega = 2 \pi \times 18^{+ 2}_{-9}~\mathrm{MHz}$ in a simple model that averages over Clebsch-Gordan coefficients. 

The classical ballistic trajectories  $\vec{r}_i(t)$ for each atom is randomly sampled from a distribution of velocities, angles, positions and arrival times that is representative of our thermal beam geometry. The atomic trajectories are generated so that atoms arrive at the cavity with a rate consistent with $N_\mathrm{eff} \approx 1$. Atoms are added or removed from the simulation when they enter or exit the y-z plane at $3w_0$ from the center of the cavity mode, where the coupling $g_i$ is roughly $0.25\%$ of its maximum possible value $g_0$.  Each atom is assigned an initial position $z_{0i}$ along the cavity axis between $0$ and $\lambda_a$ and an initial height $y_{0i}$ in the excitation laser between $-w_0$ and $w_0$. The velocity $v$ of each atom is selected from the Maxwell-Boltzmann distribution for the atomic beam flux $F(v)$, which is given by:

\begin{equation}
    F(v) = \frac{1}{2v_{rms}} \left(\frac{v}{v_{rms}}\right)^{3}e^{-\frac{v^2}{2v_{rms}^2}}
\end{equation}

\noindent where $v_{rms} = \sqrt{k_BT/m_{\mathrm{Rb}}}$ is the rms velocity of the rubidium beam, $T$ is the temperature of the atomic beam, and $m_{\mathrm{Rb}}$ is the atomic mass of $^{87}$Rb. The angular divergence of the beam is represented by a random sampling of each atom's angle from uniform distributions along the $\hat{y}$ and $\hat{z}$ directions. The distribution along the $\hat{y}$ direction takes on the full angular divergence of atomic beam $\theta_{y_{rms}} = \qty{43}{\mathrm{mrad}}$, while along $\hat{z}$ it is restricted to $\theta_{z_{rms}} = \qty{8}{\mathrm{mrad}}$. The trajectories are propagated through time as $\vec{r}_i(t) = (-3 w_0 + v_{xi}t)\hat{x}+(y_{0i} + v_{yi}t)\hat{y}+(z_{0i} + v_{zi}t)\hat{z}$.


A photon click is recorded whenever the cavity decay jump operator is applied. The generated record of time-stamped photon clicks is then supplied to our data analysis routine to generate a desired photon correlation function such as $g^{(2)}(\tau)$ or $g^{(3)}(\tau_1, \tau_2)$.

\subsection{Definitions of ${\mathrm{N}}_{\mathrm{eff}}$ and $g_{\mathrm{eff}}$}

We define an effective atom number ${\mathrm{N}}_{\mathrm{eff}}$ and effective coupling $g_{\mathrm{eff}}$ to account for both the average coupling of the atoms to the cavity and the fluctuations of the average coupling.  For $N$ atoms in the atomic beam in a cylindrical volume centered on the cavity axis $V\gg\pi w_0^2 L^2$, the spatial atomic density is just $\rho= N/V$ with Poissonian distributed atom number fluctuations in space.  However, at a specific time $t$, the atoms generate a collective vacuum Rabi splitting given by:

\begin{equation}
    g^2_\mathrm{eff} N_\mathrm{eff} (t) = \sum_{i=1}^{N} g^2_i(t)
\end{equation}

\noindent where we have expressed the time-varying coupling of the atoms to the cavity in terms of a time-independent effective coupling $g_\mathrm{eff}$ and a time-varying effective atom number $N_\mathrm{eff}(t)$.

We now consider the time averaged value $\langle\,\rangle_t$ of the above product and its variance to introduce constraints that define these two quantities.  The average can be written as

\begin{equation}
g^2_\mathrm{eff} 
\langle N_\mathrm{eff} (t)\rangle_t = \bigg{\langle}\sum_{i=1}^{N} g^2_i(t)\bigg{\rangle}_t = \int \rho\, g^2(\vec{r}) \, dV
\end{equation}

\noindent where we replace the time average by an ensemble average in the last step. 

In the limit of $V \gg \pi w_0^2 L^2$ with $\rho$ held constant, computing the variance of the coupling to the cavity $\mathrm{Var}[g^2_\mathrm{eff} N_\mathrm{eff}(t)]=g^4_\mathrm{eff} 
\langle N^2_\mathrm{eff} (t)\rangle_t  - g^4_\mathrm{eff} \langle N_\mathrm{eff}(t)\rangle_t^2 $ leads to the relationship

\begin{align*}
\mathrm{Var}[g^2_\mathrm{eff} N_\mathrm{eff}(t)] &=  \bigg{\langle}\sum_{i, j=1}^{N} g^2_i (t) \,g^2_j(t)\bigg{\rangle}_t - \bigg{\langle}\sum_{i=1}^{N} g^2_i (t)\bigg{\rangle}^2_t \\ &= \int \rho\, g^4(\vec{r}) \, dV
\end{align*}

\noindent where we have assumed no correlations in the fluctuations in coupling of atom $i$ and $j$, and again replace the time average by an ensemble average.

We now impose the final constraint that the fluctuations in the effective atom number are Poissonian such that

\begin{equation} \label{varNeff}
\mathrm{Var}[N_\mathrm{eff}(t)] = \langle N_\mathrm{eff}(t)\rangle_t
\end{equation}

\noindent from which we find that the effective coupling $g_\mathrm{eff}$ is given by

\begin{align*}
    g_{\mathrm{eff}}  &= \sqrt{\frac{\int \rho g^4(\vec{r}) dV}{\int \rho g^2(\vec{r})dV}} = \frac{1}{\sqrt{2}} \times \frac{\sqrt{3}}{2} \times g_0 \\ &= 2\pi \times \qty{16(1)}{\MHz}
\end{align*}

\noindent where the factors of $1/\sqrt{2}$ and $\sqrt{3}/2$ reflect the radial and axial contributions to the inhomogeneity, respectively. The effective atom number $N_{\mathrm{eff}}(t)$ is given by

\begin{equation}
    N_{\mathrm{eff}}(t) = \frac{8}{3}\sum_i \frac{g_i^2(t)}{g_0^2}
\end{equation}

\noindent where the sum is, in principle, taken over all atoms $i$ within some large volume $V$ in the limit $V \rightarrow \infty$. The contribution of atoms to the collective coupling rapidly decays beyond a distance $w_0$ away from the cavity axis, and the sum converges quickly to a finite value for $V\gg\pi w_0^2 L^2$. By converting the ensemble average of this expression into a time average, we can write the time-averaged effective atom number $N_\mathrm{eff} \equiv  \langle N_\mathrm{eff}(t)\rangle_t$ in terms of the atom number density $\rho$:

\begin{equation}
    N_{\mathrm{eff}} = \frac{2}{3}\pi w_0^2 L \rho
\end{equation}

This specifies an effective volume $V_{\mathrm{eff}} = \frac{2}{3}\pi w_0^2 L$, within which the mean number of atoms $\rho V_{\mathrm{eff}}$ is equal to $N_{\mathrm{eff}}$. For a flux $F$ of atoms entering the cavity at mean velocity $v$, we can write $\rho = \frac{F}{v}$ and thus

\begin{equation}
    N_{\mathrm{eff}} = V_{\mathrm{eff}}\frac{F}{v}
\end{equation}

Since $\rho$ and $F$ are not easily measurable quantities in our system, $N_{\mathrm{eff}}$ can be determined in practice from the vacuum Rabi splitting or estimated from $g^{(2)}(\tau)$ (described in the next section.)

\subsection{Bunched envelope of $g^{(2)}(\tau)$}
The bunched envelope featured in the calculated $g^{(2)}(\tau)$ disagrees with the expectation that $g^{(2)}(0) < 1$ for a single atom trapped in the cavity mode at all times. However, our atom number fluctuates as atoms enter the cavity mode at random times, and the rate of photons emitted from the cavity fluctuates in turn \cite{Foster2000, Carmichael1991}. Consider a simple model where the atoms are independent linear scatterers, which scatter photons from the excitation laser at a total rate $\Gamma_\mathrm{tot}$ proportional to $N_{\mathrm{eff}}$. In this model, we would expect:

\begin{equation}g^{(2)}(0) = \frac{\langle \Gamma_\mathrm{tot}^2\rangle}{\langle \Gamma_\mathrm{tot}\rangle^2} = 1 + \frac{\mathrm{Var}({N}_{\mathrm{eff}})}{({N}_{\mathrm{eff}})^2} = 1 + \frac{1}{N_{\mathrm{eff}}} > 1
\end{equation}

where we used the fact that $N_{\mathrm{eff}}$ is defined using equation~\ref{varNeff}. This means that $g^{(2)}(0) > 1$, which looks like photon \textit{bunching}. Note that $g^{(2)}(0)$ is larger at smaller atom number. 

This picture is complicated in reality by non-classical correlations between the emitted photons, as well as the fact that the rate of emitted photons is only exactly proportional to $N_{\mathrm{eff}}$ in the limit of small coupling $g$. Even with these complications, however, the same argument still approximately holds for our system. This phenomenon is responsible for the bunched envelope in $g^{(2)}(\tau)$.

\subsection{Heralded Correlation Measurement}
When measuring a heralded $g^{(2)}_h(\tau_a, \tau_b)$ function, we store the photon counts from all three detectors without assigning any one detector to be the herald. The choice of herald detector is arbitrary, and can be made later in post-processing. We can also arbitrarily choose which of the two remaining detectors is labeled as detector $a$ and which is labeled as detector $b$. In total, there are six possible permutations of detector labels $(a, b, c)$, and $g^{(2)}_h(\tau_a, \tau_b)$ can be calculated for any of these permutations. All possible permutations should produce the same $g^{(2)}_h(\tau_a, \tau_b)$ up to random shot noise, so we average over all permutations in (Fig.~\ref{fig4}) to reduce this noise. This introduces redundancies in the data by forcing symmetries to hold exactly (e.g. $g^{(2)}_h(\tau_a, \tau_b)$ = $g^{(2)}_h(\tau_b, \tau_a)$), but effectively increases the signal-to-noise ratio by allowing any detector to be the herald.

\section*{ACKNOWLEDGMENTS}

This material is based upon work supported by the Defense Advanced Research Projects Agency (Science of Atomic Vapors for New Technologies). Gabriela Martinez and Alexander Staron were supported under the financial assistance (Award No. 70NANB18H006) from the U.S. Department of Commerce, National Institute of Standards and Technology. The authors thank Adam Ellzey, Kim Hagen, Hans Green, and Kyle Thatcher at the JILA instrument shop for their involvement in the design and manufacturing of the presented micro-cavity, Terry Brown and James Fung-A-Fat at the JILA electronics shop for assistance in the development of electronics used to actively stabilize the cavity length, and Susan Schima for the fabrication of the microchannel array.

\end{document}